\pdfoutput=1
\documentclass{article}

\usepackage[T1]{fontenc}
\usepackage[utf8]{inputenc}
\usepackage{lmodern}
\usepackage[polish,english]{babel}
\usepackage{float}

\usepackage{graphicx}
\usepackage{multirow}
\usepackage{amsmath,amssymb,amsfonts}
\usepackage{amsthm}
\usepackage{mathrsfs}
\usepackage[title]{appendix}
\usepackage{xcolor}
\usepackage{textcomp}
\usepackage{manyfoot}
\usepackage{booktabs}
\usepackage{algorithm}
\usepackage{algorithmicx}
\usepackage{algpseudocode}
\usepackage{listings}
\usepackage{geometry}
\usepackage{tabularx}
\usepackage{microtype}

\theoremstyle{thmstyleone}

\theoremstyle{thmstyletwo}

\theoremstyle{thmstylethree}

\newcommand{\keywords}[1]{\noindent\textbf{Keywords: }#1}
\usepackage[numbers,sort&compress]{natbib}

\newenvironment{highlights}{\section*{Highlights}\begin{itemize}}{\end{itemize}}

\usepackage[hidelinks,unicode]{hyperref}

\title{Impact of Neuron Models on Spiking Neural Networks performance. A Complexity Based Classification Approach}

\author{
  Zofia Rudnicka \and
  Janusz Szczepanski \and
  Agnieszka Pregowska\thanks{Corresponding author: \texttt{aprego@ippt.pan.pl}}
}

\date{Institute of Fundamental Technological Research,\\
Polish Academy of Sciences,\\
Pawinskiego 5B, 02-106 Warsaw, Poland}

\begin{document}
\maketitle
%% Abstract
\begin{abstract}
This study explores how the selection of neuron models and learning rules impacts the classification performance of Spiking Neural Networks (SNNs), with a focus on applications in bio-signal processing. We compare biologically inspired neuron models, including Leaky Integrate-and-Fire (LIF), metaneurons, and probabilistic Levy-Baxter (LB) neurons, across multiple learning rules, including spike-timing-dependent plasticity (STDP), tempotron, and reward-modulated updates. A novel element of this work is the integration of a complexity-based decision mechanism into the evaluation pipeline. Using Lempel-Ziv Complexity (LZC), a measure related to entropy rate, we quantify the structural regularity of spike trains and assess classification outcomes in a consistent and interpretable manner across different SNN configurations. To investigate neural dynamics and assess algorithm performance, we employed synthetic datasets with varying temporal dependencies and stochasticity levels. These included Markov and Poisson processes, well-established models to simulate neuronal spike trains and capture the stochastic firing behavior of biological neurons.Validation of synthetic Poisson and Markov-modeled data reveals clear performance trends: classification accuracy depends on the interaction between neuron model, network size, and learning rule, with the LZC-based evaluation highlighting configurations that remain robust to weak or noisy signals. This work delivers a systematic analysis of how neuron model selection interacts with network parameters and learning strategies, supported by a novel complexity-based evaluation approach that offers a consistent benchmark for SNN performance.
\end{abstract}

\begin{highlights}
\item First systematic study of neuron model impact on SNN performance
\item Compared LIF, metaneurons, and Levy-Baxter across varied network sizes
\item Evaluated supervised, unsupervised, and hybrid learning rules
\item Introduced LZC-based decision metric for SNN classification outputs
\item LZC identifies best neuron versus learning rules for biosignal tasks
\end{highlights}

%% Keywords

\section{Introduction}
\label{intro}
Neural communication is a highly complicated and dynamic process through which neurons convey information within biological systems. The fundamental mechanism of this communication involves the propagation of electrical signals, known as action potentials \cite{Gerstner2014}. Artificial Neural Networks (ANNs) are designed to emulate aspects of this biological communication process through a system of interconnected computational nodes, or "neurons." These nodes process information via mathematical operations, and their interconnections are modifiable based on learning algorithms, mirroring the adaptive mechanisms observed in biological neural networks. The connections in neural networks are adjusted based on feedback, allowing them to “learn” from data and optimize their performance, similar to the synaptic modifications in the brain. ANNs utilize simplified mathematical models that simulate the underlying processes of neural communication. These models operate within a hidden layer, where they are connected to the output layer of the network. The computations within these nodes are based on mathematical operations, and their output is combined with the weights of the connections between nodes, which are adjusted during training, mainly with algorithms such as backpropagation \cite{Lin2016}. In this framework, the training process involves iteratively adjusting the weights of connections based on the discrepancy between the predicted outputs and the actual outputs. This adjustment allows the network to "learn" and improve its performance over time. Although artificial neurons are designed to mimic certain aspects of biological neurons, they do so in a much more abstract and simplified manner. The complexity of biological neurons, influenced by a wide range of internal and external factors, is reduced to a basic computational model that focuses on input-output relationships.
\par
While both biological and artificial neurons process information, the mechanisms and complexity differ significantly \cite{Seguin2023}. Biological neurons operate in highly dynamic environments, with their activity influenced by a multitude of biochemical processes and external stimuli. In contrast, artificial neurons, governed by specific architectures and learning algorithms, are simplified representations of this complexity. This difference highlights the inherent contrast between the adaptive, biochemical complexity of biological systems and the structured, algorithmic framework of computational models.  Thus, the limitations of classical Artificial Intelligence, particularly in models based on perceptrons and traditional ANNs naturally forced the exploration of alternative neural network architectures that can improve computational efficiency. One of the most promising candidates for overcoming these limitations seems to be Spiking Neural Networks (SNNs), which are considered to be a more energy-efficient option for complex calculations \cite{datta2022}. The key distinction between SNNs and conventional ANNs lies in their output dynamics. Unlike ANNs, SNNs utilize a spiking mechanism, where information is transmitted as discrete temporal spikes rather than as continuous signals. This dynamic spike-based signaling more closely emulates the way information flows through biological synapses, allowing SNNs to represent features in spatiotemporal data more effectively. Consequently, SNNs exhibit a promising ability to perform computations in a manner that approaches the temporal processing capabilities of the human brain, enabling richer and more efficient representations of time-dependent data \cite{Sun2025}.
\par
When studying the efficiency of information transmission in SNN, the selection of both neuron and network architecture models is crucial. A model that accurately replicates the spike response of a neuron to any input current is fundamental for both constructing brain simulators and understanding the computational mechanisms of neural activity \cite{Izhikevich2003,Shirsavar2023}. Several approaches to neuron modeling are being developed \cite{Gerstner2014}, with two primary lines of development being most prominent. The first approach involves detailed biophysical modeling, such as Hodgkin and Huxley-like models, which describe the dynamics of ion channels within the spatially structured tree-like morphology of neurons. The second approach includes the integrate-and-fire (LIF) models \cite{Dutta2017}, which treat neuronal electrical activity as a threshold-based process. As network building blocks (single-neuron models), first, we will assume perceptorns, than recently used in SNNs, Leaky Integrate-and-Fire neuron model, metaneurons \cite{weng2021} and probabilistic Levy Baxter neuron model \cite{LevyBaxter2002, Paprocki2020}, which provide results consistent with physiological observed values. In this paper, we analyze how the performance of the SNN is influenced by the neuron model used to build it. 
We introduce a novel hybrid framework that integrates the temporal precision and biological plausibility of SNNs with the Lempel-Ziv complexity (LZC) measure \cite{LempelZiv1976} to improve the classification of spatiotemporal neural data. By quantifying the structural complexity of spike patterns, the proposed method offers interpretable and noise-robust classification, particularly effective for data exhibiting variable temporal dynamics, such as Poisson-distributed signals.

\section{Main Contribution}
The main contributions of the paper are as follows:
\begin{itemize}
\item Comprehensive analysis of how different neuron models: Leaky Integrate-and-Fire (LIF), metaneurons, and probabilistic Levy-Baxter, affect SNN classification performance across varying network sizes and tasks.
\item Systematic evaluation of multiple learning algorithms, including unsupervised (STDP, SDSP), supervised (tempotron, backpropagation), and hybrid reward-modulated approaches, to quantify their interaction with neuron model choice.
\item Demonstration of effectiveness of bio-inspired neuron models in bio-signal classification, achieving high accuracy, sensitivity, and specificity on synthetic datasets modeled with Poisson and Markov processes.
\item Introduction of a complexity-based evaluation by integrating Shannon’s Information Theory with SNN outputs, using Lempel–Ziv Complexity to capture subtle temporal structure in spike trains and improve detection of weak or noisy signals.
\item Identification of performance trends showing how optimal neuron model–learning rule combinations depend on signal characteristics and network size, with tempotron and reward-based learning showing notable advantages in certain regimes.
\item Highlighting the potential of SNNs as a biologically plausible and computationally efficient framework for processing complex spatio-temporal data, particularly biosignals.
\end{itemize}

\section{Basics notation}
In Table \ref{tab:1} the notation used is presented. 
\begin{table}[H]
    \centering
    \begin{tabular}{|c|c|}
    \hline
    \textbf{Description} & \textbf{Notation} \\
    \hline
    input vector  & $\textbf{x}=[x_{1},x_{2},\cdots,x_{n}] \in \mathbb{R}^{n}$ \\
    weight vector & $\textbf{w}=[w_{1},w_{2},\cdots,w_{n}] \in \mathbb{R}^{n}$  \\
    bias & $b$ \\
    threshold & $\theta$ \\
    weighted sum & $z=\textbf{w}^{T} \textbf{x} + b \in \mathbb{R}$\\
    activation function  & $f(z), f: \mathbb{R} \rightarrow  \mathbb{R}$ or $f: \mathbb{R} \rightarrow \{0,1\}$\\
    \hline
    \end{tabular}
    \caption{Basic notation used in the study.}
    \label{tab:1}
\end{table}

\section{Models of Neurons}
\label{models}
The perceptron operates in an $n$-dimensional real vector space $\mathbb{R}^{n}$. The input vector is $\textbf{x} = [x_1, x_2, \dots, x_n]$, and the weight vector is $\textbf{w} = [w_1, w_2, \dots, w_n] \in \mathbb{R}^n$. The perceptron computes a weighted sum of inputs with an added bias $b$, as given by:
\begin{equation}
    z = \sum_{i=1}^n w_i x_i + b
    \label{perce}
\end{equation}
where $b \in \mathbb{R}$ is the bias term, and the output is determined by a threshold function $f(z)$:
\begin{equation}
    f(z) = \begin{cases}
    1 & \text{if } z \geq \theta \\
    0 & \text{otherwise}
    \end{cases}
    \label{step}
\end{equation}
where $\theta$ is the activation threshold \cite{Parlos1994}.
\par
The Leaky Integrate-and-Fire (LIF) model describes the membrane potential dynamics of a neuron \cite{Dutta2017}. The membrane potential $U(t)$ evolves over time according to the equation:
\begin{equation}
    \tau_m \frac{dU}{dt} = -U(t) + R_m I(t)
    \label{LIFeq}
\end{equation}
where $\tau_m$ is the membrane time constant, $R_m$ is the membrane resistance, and $I(t)$ is the input current at time $t$. When the membrane potential exceeds a threshold $U_{th}$, a spike occurs, and the potential is reset to a lower value $U_r$. This process models the gradual accumulation and leakage of membrane potential.
\par
The metaneuron is a higher-level computational unit that abstracts the activity of multiple neurons or neural processes \cite{Cheng2023}. It introduces a modular approach to neural network design, facilitating large-scale modeling by representing groups of neurons or collective behaviors. Like the perceptron, it computes a weighted sum of inputs, but with greater flexibility in activation functions, supporting binary step, sigmoid, or ReLU. Unlike the perceptron, the metaneuron can model dynamic neuron behaviors, including spiking dynamics, by processing time-varying inputs and evolving internal states, akin to LIF model. This generalization allows it to capture complex neural phenomena such as oscillations and synchrony, making it highly adaptable for large-scale networks.
\par
The Levy-Baxter model incorporates probabilistic dynamics to capture synaptic transmission variability \cite{LevyBaxter2002, Paprocki2020}. The inputs to the neuron are represented by the vector $\textbf{x} = [x_1, x_2, \dots, x_n]$, with each $x_i$ modeled as a binary stochastic process. The transformation of each input is governed by a Bernoulli random variable $\phi_i$ with success probability $s$, and the amplitude is scaled by a random variable $Q_i$ uniformly distributed over $[0, 1]$. The transformed input is given by:
\begin{equation}
    \textbf{y} = [\phi_1 Q_1 x_1, \phi_2 Q_2 x_2, \dots, \phi_n Q_n x_n]
\end{equation}
The total excitation $\sigma$ is the sum of the transformed inputs:
\begin{equation}
    \sigma = \sum_{i=1}^n \phi_i Q_i x_i
\end{equation}
The output of the neuron is determined by a threshold function $f(\sigma)$:

\begin{equation}
    z = \begin{cases}
    1 & \text{if } \sigma \geq 0 \\
    0 & \text{if } \sigma < 0
    \end{cases}
    \label{LBstep}
\end{equation}

where $z = 1$ indicates the neuron has fired, and $z = 0$ indicates no spike.
This probabilistic approach models a synaptic variability, with $x_i$ as inputs, $\phi_i$ as quantal release probabilities, and $Q_i$ as the scaling factor representing amplitude fluctuations.
\par
Each neuron model described here represents a different abstraction of neural behavior. The perceptron is a simple threshold-based model used for binary classification tasks, operating deterministically. The spiking phenomenon is not included. The LIF model incorporates temporal dynamics, modeling the gradual integration of inputs and natural leakage of membrane potential. It introduces time-dependent behavior and spiking. The metaneuron abstracts the collective activity of multiple neurons into a higher-level computational unit, enabling modular and hierarchical network design. The spiking phenomenon is optional. The L-B model introduces stochastic and quantal variability in synaptic transmission, providing a probabilistic framework for understanding neural responses. While stochastic provides more realistic noise modeling.

\section{Spiking Neural Network Architecture}
Spiking Neural Networks process information by considering spike signals, making them particularly promising for handling complex tasks \cite{
Huang2022}. These networks excel at encoding intricate spatiotemporal information through spike patterns. The design of SNNs is often based on models like LIF neuron model, which is a simplified representation of how biological neurons process information. In these models, the arrival of a spike at a presynaptic neuron triggers an input current $I(t)$ that influences the membrane potential of the postsynaptic neuron. For simplicity, we can express the input current as a convolution of the spike signal $S_{j}(t)$ rom a presynaptic neuron with an exponential decay function, representing the temporal filtering of the spike signal
\begin{equation}
    I(t)= \int_{0}^{\infty} S_{j}(s-t) exp( \frac{-s}{\tau_{s}}) ds, 
\end{equation}
where $S_{j}(s-t)$ represents the spike train from the $j$-th presynaptic neuron, and $\tau_{s}$ is a synaptic time constant, dictating the decay of the signal over time. This equation models the temporal dynamics of the input signal, which integrates over time, decaying with rate  $\tau_{s}$.

\begin{figure}
    \centering
    \includegraphics[width=0.8\linewidth]{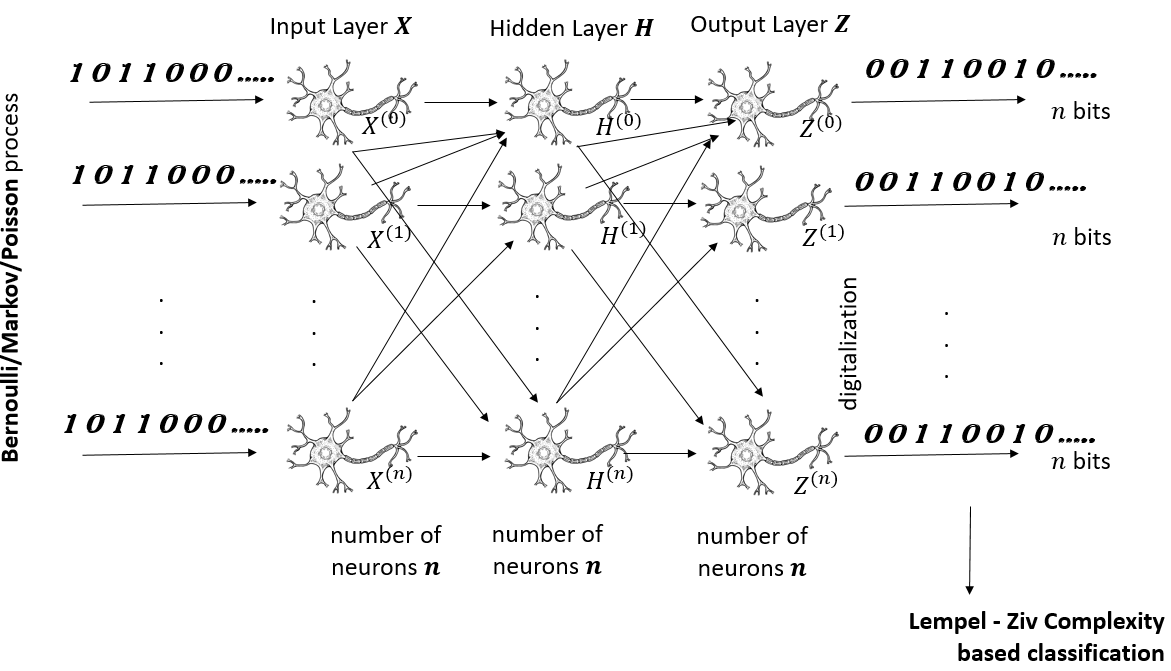}
    \caption{The basics scheme of the conducted classification task.}
    \label{fig:1}
\end{figure}

The neural network consists of three layers of neurons: input, hidden, and output, each containing \(n\) neurons, as shown in Figure \ref{fig:1}. We considered neural networks comprising 16, 32, 64, 128, 256, 512, and 1024 neurons per layers. The best of the results obtained are presented in Table 1, Table 2, Table 3, and Table 4 and Table 5. 
Sequences of binary values (strings of zeros and ones), each 1024 bits long, are fed into the network. Then, \(n\)-bit-long sequences of action potentials (spike trains) are generated by the network and subsequently converted back into sequences of zeros and ones. 
\par
Shannon's Information Theory establishes a mathematical framework for quantifying and analyzing the transmission of information within communication systems \cite{Shannon1948}. Thus, entropy rate estimators provide a rigorous mathematical approach to approximating information transmission rates, offering an alternative to traditional firing rate analysis. Notably, Lempel–Ziv complexity, as defined by Lempel and Ziv (1976) \cite{LempelZiv1976}, has been successfully employed as an effective estimator. The Lempel-Ziv complexity is a widely used metric for estimating entropy, and consequently the informational content of sequences. Given a sequence \( \textbf{x}_n^1 := [x_1, x_2, \dots, x_n] \), where each \( x_i \) belongs to a finite source alphabet (e.g., \( x_i \in \{0, 1\} \)), the complexity \( C_{\alpha}(\textbf{x}_n^1) \) counts the number of distinct blocks (or patterns) in the sequence. A new block is defined when a substring starting from the current position has not appeared before. The normalized complexity, \( c_{\alpha}(\textbf{x}_n^1) \), which measures the rate at which new patterns are generated, is defined as:
\begin{equation} \label{lzc_from}
c_{\alpha}(\mathbf{x}_n^1) = \frac{C_{\alpha}(\mathbf{x}_n^1)}{n}\log_{\alpha}{n},
\end{equation}
where  $\alpha = 2 $ for binary sequences. Asymptotically, $ c_{2}(\mathbf{x}_n^1) \to 1 $ for random sequences and $ c_{2}(\mathbf{x}_n^1) \to 0 $ for deterministic sequences. For random sequences, the normalized complexity tends to 1, and for deterministic sequences, it approaches 0. The LZ complexity serves as an effective estimator of entropy for ergodic stochastic processes \cite{LempelZiv1976, Arnold2013, Pregowska2016,Pregowska2019}. In neural networks, Lempel-Ziv complexity is applied to classify output sequences based on their informational content and unpredictability. The output sequences are classified using the Lempel-Ziv 1976 complexity-based classifier.
Various neuron models were used to construct the SNN, including the perceptron, LIF, metaneuron, and LB neuron models. Moreover, different types of learning algorithms, including unsupervised methods (e.g., spike-timing dependent plasticity STDP \cite{Kheradpisheh2016}), spike-driven synaptic plasticity SDSP \cite{Rahman2025}), supervised methods (e.g., tempotron \cite{Gutig2006,Yu2014}, backpropagation \cite{Wu2018}), and hybrid methods (e.g., reward-based learning with active learning \cite{Zhan2023}), have been widely investigated to optimize network performance.

\section{Input datasets}
Neuronal action potentials and spike sequences are often modeled as point processes, providing a statistical framework for analyzing discrete events over time \cite{DaleyVereJones2003,Wojcik2009}. Poisson point processes are particularly useful for representing spike trains with minimal temporal dependencies, effectively capturing the stochastic nature of neuronal firing \cite{Rieke1997}. Their Markov properties make them suitable for modeling short-term memoryless behavior in neural dynamics \cite{PapoulisPillai2002}.

\par
These insights emphasize the need for probabilistic and temporally structured input data to evaluate the capacity of computational models to represent neural signals accurately. To this study uses three types of synthetic binary sequences as input datasets: Bernoulli sequences, first-order Markov sequences, and Poisson-based spike trains. Each of these allows for controlled manipulation of randomness, temporal dependence, and rate-based variability, i.e. factors critical for evaluating the interplay between neural dynamics and classification mechanisms based on complexity.

\par
Bernoulli sequences are generated from independent Bernoulli processes, where each binary element \( B_i \) is drawn independently with probability \( p \). For our experiments, we generate two sets of binary sequences \( B_1 \) and \( B_2 \), each of length 1024, using different values of \( p \). The independence of events allows us to assess the system's performance under random binary outcomes with varying probabilities.

\par
Markov sequences are generated from a first-order Markov process, where the probability of each element \( M_i \) depends on the previous state \( M_{i-1} \). For our experiments, two sets of binary sequences \( M_1 = \{m_{1i}\}, i = 1, 2, \dots, 1000 \) and \( M_2 = \{m_{2i}\}, i = 1, 2, \dots, 1000 \) are generated with transition probabilities defining the dependency structure. We explore different transition probability configurations to evaluate the impact of state dependencies on classification accuracy.

\par
Poisson sequences are generated from a Poisson process, where events occur independently at a constant average rate \( \lambda \). For each Poisson process, spike trains are generated with rate parameters \( \lambda_x \) and \( \lambda_y \). These sequences are tested under various rate configurations to observe the effect of spike timing variability on system behavior and classification performance.
\par
To explore neural dynamics, we use Poisson and Markov processes to model neuronal spike trains, a well-established method for simulating the stochastic firing behavior of biological neurons.

\par

The rationale for using synthetic sequences with controlled stochastic properties is to enable a rigorous evaluation of how SNNs and biologically inspired learning algorithms respond to different temporal structures and statistical dependencies. Unlike real-world benchmarks, these synthetic inputs allow for precise manipulation of noise, memory, and event distributions,  i.e. factors that are crucial for understanding the encoding capabilities of SNNs. Moreover, the use of interpretable, parameterized inputs facilitates a clearer attribution of classification outcomes to underlying neural dynamics and complexity-based decision mechanisms.

\section{Related works}
In the paper by \cite{Dan2025}, a Spiking Neural Network architecture based on LIF neuron model was proposed. The authors introduced a residual-based SNN architecture with dynamic threshold adjustment, which combines direct encoding (frequency-based neuronal representation) with multineuronal population decoding. The MNIST dataset was considered, and the architecture achieved notable performance in just 6 time-steps, with accuracy improvements ranging between 1.00\% and 7.50\%, depending on the dataset. Similarly, \cite{Luo2025} proposed an SNN based on a current-based adaptive LIF (CuAdLIF) neuron model featuring delayed responses and membrane potential adaptation. This design improves temporal correlations and maintains long short-term memory. On the other hand, \cite{Zhao2025} advanced neuromorphic computing by developing a high-performance neuromorphic processing unit (NPU) tailored for high data throughput and robust SNN processing. This NPU utilized LIF neurons with a backpropagation-based spike-timing-dependent plasticity learning algorithm, achieving an accuracy of 91.00\% on the MNIST dataset. This approach set a new benchmark in neuromorphic computing, offering superior data throughput and neural processing precision compared to systems like SpiNNaker 2.
\par
Study \cite{Liu2025} applied the STDP learning algorithm to an SNN based on LIF neurons for classification tasks on the MNIST dataset, showing promising results and highlighting the potential of this research direction. Another bio-inspired learning algorithm tested in SNNs is the tempotron, as reported by \cite{Yu2014}, who achieved high classification accuracy on the MNIST dataset. Building on this, \cite{Patankar2025} employed a tempotron learning rate method for training and standard STDP for optimization in an SNN based on LIF neurons for medical data applications. This approach significantly improved processing speed and reduced complexity compared to other SNN methods. Various cross-validation techniques were used to validate the robustness of the model, demonstrating its superiority over existing state-of-the-art SNNs. Additionally, \cite{Luo2025} addressed the limitations of SNNs based on meta neuron models by improving the You-Only-Look-Once (YOLO) algorithm. Their SpikeYOLO architecture, i.e. a simplified version of YOLO incorporating meta SNN blocks, which minimized spike degradation and improved detection accuracy.
\par
Despite these advances, none of the aforementioned studies considered the impact of neuron models on SNN performance, particularly the effects on accuracy and computation time in relation to the number of neurons per layer and different neural network learning algorithms. All the reviewed works relied on the LIF neuron model or its variations, such as the meta neuron, and there is a clear trend toward implementing biologically inspired algorithms in SNNs based on LIF neurons. However, these studies did not investigate the influence of network size or the potential of alternative neuron models. Moreover, to our knowledge, no existing research has explored combining SNNs with concepts derived from Information Theory to enhance network accuracy while reducing computation time. This gap in the literature underscores the novelty of our proposed approach.

\section{Results}
\par
The considered datasets were divided into two subsets: 90.00\% for training and 10.00\% for testing, ensuring that the models had sufficient data to learn while retaining a separate evaluation set. Accuracy was used the primary metric for evaluating classification performance. All computations were provided on Intel(R) Core(TM) i7-14700F, 2.10 GHz.
\par
First, we consider four spiking neural networks made of LIF, perceptron, meta neurons, LB neuron model as well as the hybrid network, in which input layer was construed by LIF neurons, while the hidden and output layers were made of perceprtrons, respectively. In Table \ref{table:combined} the influence of neuron model on commonly used learning algorithm, i.e. BP learning algorithm was presented. It turned out that all cases gave accuracies above 90.00\%, except for the use of a network composed of the LIF model to the Poisson source, however, the computation times differed significantly.
The application of the compared to a neural network composed of LIF neurons, the use of the meta nueron model in the case of BP learning algortihms gives higher accuracy in comparable computation time. Surprisingly, for input data in the form of the Bernoulli process (actually the simplest data set), a neural network consisting of 512 LIF neuron in input layer and 512 perceptrons in hidden and output layers was needed to achieve high accuracy. In comparison, architectures composed of the remaining neurons models required only 64 neurons per layer. Moreover, in the case of the Poisson process, the network based on LIF neurons also required four times more epochs to achieve lower accuracy than in other cases.

\par
Table \ref{table:combined_2} shows the influence of neuron model on bio-inspired learning algorithms like tempotron learning rule, BAL, STDP, and SDSP. We consider the same architectures as in Table \ref{table:combined}. The results obtained show that involving biologically inspired learning algorithms in the process of training a neural network allows for significantly shortened computation time, especially when the neural network is built from LIF, meta, and LB neuron models. The meta and LB neurons models consistently demonstrated high efficiency in handling Bernoulli processes across different bio-inspired learning algorithms. It achieved a balance between accuracy and computational time, making it well-suited for time-sensitive applications requiring moderate accuracy. The first approach demonstrates high efficacy for Bernoulli sequences, where the independence between events reduces the necessity for complex temporal integration. The Levy-Baxter model, although slightly slower, consistently achieved perfect accuracy across all Bernoulli datasets. The introduction of perceptrons into neural networks presented trade-offs: while they occasionally improved processing speed, particularly in the BAL and tempotron algorithms, they generally led to reduced accuracy and significantly increased training times, as observed in the case of STDP-based algorithms. The Levy-Baxter neural model exhibited superior accuracy, consistently reaching 99.00-100.00\% in all scenarios tested. Despite requiring slightly longer training times compared to simpler neuron models, its adaptability and robustness in BAL scenarios make it a strong candidate for accuracy-critical applications. The application of the tempotron learning algorithm enables high-accuracy computations within optimal time constraints. Furthermore, meta-neurons demonstrated an accuracy range of 90 to 100\% when implemented in SNNs with 32 to 64 neurons, while other neuron models typically required 128 neurons or more to achieve comparable performance.
\par
The analysis of neuron models under BAL, tempotron, and STDP learning algorithms for Markov sequences revealed distinct performance trends, emphasizing the trade-offs between accuracy and computational efficiency. The perceptron model consistently excelled, achieving perfect accuracy (i.e., 100.00\%) in BP and STDP scenarios, demonstrating its effectiveness in capturing Markovian dependencies. Hybrid models integrating LIF neurons with perceptrons exhibited significant potential under BP and bio-inspired learning algorithms, balancing low computational cost with high accuracy. However, these models showed inefficiencies under SDPD, where accuracy declined to 83.00\%. The biologically inspired LB model demonstrated exceptional robustness, achieving near-perfect accuracy (93.75–100.00\%) across tasks. However, its substantially longer runtimes make it more suitable for precision-critical applications rather than time-sensitive computations. Conversely, meta neurons provided a compelling balance between accuracy and efficiency, maintaining high accuracy while requiring only 32-64 neurons per layer, significantly reducing computational overhead compared to alternative neuron models.
\par
For datasets containing Poisson processes, the perceptron-based neural network model consistently demonstrated the highest efficiency and effectiveness, achieving perfect accuracy (i.e. 100.00\%) in minimal time under the tempotron learning algorithm and SDPS. However, in the case of BAL, the computational time increased by an order of magnitude while maintaining the same accuracy. 
Hybrid models, such as those combining LIF neurons with perceptrons, exhibited significantly lower accuracy, ranging from 68.00\% to 89.00\%, but with relatively low computational costs. While this approach benefits from step dynamics and linear decision boundaries, it requires careful tuning to prevent inefficiencies, particularly with increased learning periods or larger network sizes. Under the SDSP algorithm, a configuration with 16 neurons and 10 epochs achieved an accuracy of 68.00\% with a runtime of 28.1 seconds. Similarly, under the STDP algorithm, the same configuration yielded the same accuracy (i.e. 68.00\%) but with a significantly reduced runtime of 2.7 seconds, highlighting the computational efficiency of STDP. In contrast, applying the BAL algorithm to a larger network configuration (64 neurons, 20 epochs) resulted in a significantly higher accuracy of 88.00\%, with a runtime of 1 minute and 48.0 seconds. Biologically inspired models, such as the LB model, exhibited robustness and representational richness, achieving accuracy levels between 91.00\% and 99.00\% with 64–128 neurons under bio-inspired learning algorithms, albeit at the cost of increased computational time. In turn, metaneuron networks consistently provided high accuracy with low computational costs for 64 neurons per layer across all bio-inspired learning algorithms, except for tempotron. In the case of the tempotron learning algorithm, achieving comparable accuracy required 128 neurons per layer. Nevertheless, the computational cost remained an order of magnitude lower than in previous cases, reinforcing the efficiency of metaneuron networks.

\par
The optimal neuron model and learning algorithm depend on the application’s accuracy and efficiency requirements. LB models excel in accuracy-critical tasks, while LIF-based architectures with BAL provide efficient solutions. Hybrid models offer a promising middle ground, performing well when paired with appropriate learning algorithms.

% w preambule dodaj (jeśli nie masz):
% \usepackage{float}

\setcounter{table}{0}
\begin{table}[H]
\centering
\caption{Influence of neuron model on commonly used learning algorithm. The BP learning algorithm was applied.}
\label{table:combined}
\resizebox{0.9\linewidth}{!}{%
\begin{tabular}{|p{3cm}|p{6cm}|p{4cm}|p{3cm}|p{3cm}|p{3cm}|}
\hline
\textbf{Input data} & \textbf{Neuron model} & \textbf{Neurons number in each layer} & \textbf{Epochs number} & \textbf{Time} & \textbf{Accuracy [\%]} \\[0.5ex]
\hline\hline

Bernoulli & LIF & 128 & 10 & 12m36.8s & 99.00 \\
Bernoulli & LIF in input layer + perceptron in hidden and output layers & 512 & 20 & 2m26.7s & 100.00 \\
Bernoulli & Perceptron & 128 & 10 & 17.0s & 100.00 \\
Bernoulli & Meta & 32 & 10 & 7.8s & 99.50 \\
Bernoulli & Levy-Baxter & 128 & 20 & 11m58.6s & 100.00 \\
\hline\hline

Markov & LIF & 128 & 10 & 12m36.8s & 97.64 \\
Markov & LIF in input layer + perceptron in hidden and output layers & 64 & 20 & 2m26.7s & 100.00 \\
Markov & Perceptron & 32 & 20 & 17.0s & 93.94 \\
Markov & Meta & 32 & 10 & 1m7.4s & 98.75 \\
Markov & Levy-Baxter & 64 & 20 & 11m58.6s & 100.00 \\
\hline\hline

Poisson & LIF & 64 & 40 & 12m36.8s & 87.00 \\
Poisson & LIF in input layer + perceptron in hidden and output layers & 64 & 20 & 2m26.7s & 99.50 \\
Poisson & Perceptron & 64 & 10 & 17.0s & 100.00 \\
Poisson & Meta & 32 & 10 & 40.7s & 96.00 \\
Poisson & Levy-Baxter & 64 & 10 & 11m58.6s & 97.00 \\
\hline
\end{tabular}%
}
\end{table}

\begin{table}[H]
\centering
\caption{Influence of neuron model on bio-inspired learning algorithms.}
\label{table:combined_2}
\resizebox{0.9\linewidth}{!}{
\begin{tabular}{|p{3cm}|p{6cm}|p{3cm}|p{4cm}|p{4cm}|p{3cm}|p{3cm}|}
\hline
\textbf{Input data} & \textbf{Neuron model} & \textbf{Learning algorithm}  & \textbf{Neurons number in each layer} & \textbf{Epochs number} & \textbf{Time} & \textbf{Accuracy [\%]} \\  
\hline\hline

Bernoulli & LIF & tempotron & 128 & 10 & 7.02s & 96.00 \\
Bernoulli & LIF in input layer + perceptron in hidden and output layers & tempotron & 128 & 10 & 19.9s & 100.00 \\
Bernoulli & Perceptron & tempotron & 128 & 10 & 19.6s & 100.00 \\
Bernoulli & Meta & tempotron & 64 & 40 & 1m24.4s & 90.50 \\
Bernoulli & Levy-Baxter & tempotron & 128 & 10 & 2m38.4s & 99.50 \\
\hline
Bernoulli & LIF & SDSP & 128 & 45 & 3m2.3s & 89.00 \\
Bernoulli & LIF in input layer + perceptron in hidden and output layers & SDSP & 128 & 10 & 12m1.7s & 87.50 \\
Bernoulli & Perceptron & SDSP & 128 & 10 & 17.0s & 100.00 \\
Bernoulli & Meta & SDSP & 32 & 10 & 3m56.0s & 100.00 \\
Bernoulli & Levy-Baxter & SDSP & 128 & 10 & 6m48.0s & 99.00 \\
\hline
Bernoulli & LIF & STDP & 128 & 10 & 3m44.4s & 91.50 \\
Bernoulli & LIF in input layer + perceptron in hidden and output layers & STDP & 128 & 10 & 11m50.9s & 87.50 \\
Bernoulli & Perceptron & STDP & 512 & 10 & 86m48.8s & 100.00 \\
Bernoulli & Meta & STDP & 32 & 10 & 20.2s & 92.50 \\
Bernoulli & Levy-Baxter & STDP & 128 & 10 & 3m49.0s & 100.00 \\
\hline
Bernoulli & LIF & BAL & 128 & 30 & 3m28.9s & 86.50 \\
Bernoulli & LIF in input layer + perceptron in hidden and output layers & BAL & 128 & 10 & 4.7s & 87.50 \\
Bernoulli & Perceptron & BAL & 128 & 10 & 5m38.5s & 99.00 \\
Bernoulli & Meta & BAL & 32 & 10 & 6.7s & 99.50 \\
Bernoulli & Levy-Baxter & BAL & 32 & 10 & 22.2s & 100.00 \\
\hline\hline

Markov & LIF & tempotron & 128 & 10 & 40.5s & 90.59 \\
Markov & LIF in input layer + perceptron in hidden and output layers & tempotron & 64 & 10 & 29.1s & 96.75 \\
Markov & Perceptron & tempotron & 128 & 10 & 53.6s & 100.00 \\
Markov & Meta & tempotron & 64 & 10 & 1m8.2s & 92.00 \\
Markov & Levy-Baxter & tempotron & 64 & 10 & 36.0s & 93.75 \\
\hline
Markov & LIF & SDSP & 128 & 10 & 1m46.4s & 92.25 \\
Markov & LIF in input layer + perceptron in hidden and output layers & SDSP & 256 & 10 & 93m2.4s & 83.00 \\
Markov & Perceptron & SDSP & 128 & 20 & 3m44.7s & 100.00 \\
Markov & Meta & SDSP & 64 & 10 & 53.1s & 97.53 \\
Markov & Levy-Baxter & SDSP & 128 & 10 & 11m9.2s & 95.50 \\
\hline
Markov & LIF & STDP & 128 & 10 & 3m45.1s & 91.50 \\
Markov & LIF in input layer + perceptron in hidden and output layers & STDP & 64 & 10 & 1m17.3s & 85.00 \\
Markov & Perceptron & STDP & 512 & 10 & 195m34.2s & 100.00 \\
Markov & Meta & STDP & 64 & 10 & 51.3s & 99.20 \\
Markov & Levy-Baxter & STDP & 128 & 10 & 1m47.9s & 80.75 \\
\hline
Markov & LIF & BAL & 128 & 10 & 1m56.2s & 96.25 \\
Markov & LIF in input layer + perceptron in hidden and output layers & BAL & 256 & 20 & 11m12.9s & 95.00 \\
Markov & Perceptron & BAL & 512 & 10 & 266m55.1s & 99.75 \\
Markov & Meta & BAL & 32 & 10 & 7.6s & 99.50 \\
Markov & Levy-Baxter & BAL & 64 & 10 & 2m12.4s & 96.25 \\
\hline\hline

Poisson & LIF & tempotron & 128 & 10 & 7.3s & 89.50 \\
Poisson & LIF in input layer + perceptron in hidden and output layers & tempotron & 128 & 20 & 3m23.9s & 89.00 \\
Poisson & Perceptron & tempotron & 64 & 10 & 8.5s & 100.00 \\
Poisson & Meta & tempotron & 128 & 10 & 7.8s & 89.50 \\
Poisson & Levy-Baxter & tempotron & 64 & 10 & 5.8s & 91.00 \\
\hline
Poisson & LIF & SDSP & 64 & 10 & 32.2s & 90.00 \\
Poisson & LIF in input layer + perceptron in hidden and output layers & SDSP & 16 & 10 & 28.1s & 68.00 \\
Poisson & Perceptron & SDSP & 16 & 10 & 3.1s & 100.00 \\
Poisson & Meta & SDSP & 64 & 10 & 26.1s & 91.00 \\
Poisson & Levy-Baxter & SDSP & 128 & 10 & 20m57.7s & 99.00 \\
\hline
Poisson & LIF & STDP & 128 & 20 & 10m44.7s & 97.50 \\
Poisson & LIF in input layer + perceptron in hidden and output layers & STDP & 16 & 10 & 2.7s & 68.00 \\
Poisson & Perceptron & STDP & 64 & 10 & 10.7s & 91.00 \\
Poisson & Meta & STDP & 64 & 10 & 31.9s & 88.00 \\
Poisson & Levy-Baxter & STDP & 128 & 10 & 8m7.9s & 95.00 \\
\hline
Poisson & LIF & BAL & 128 & 30 & 10m6.9s & 96.00 \\
Poisson & LIF in input layer + perceptron in hidden and output layers & BAL & 64 & 20 & 1m48.0s & 88.00 \\
Poisson & Perceptron & BAL & 64 & 10 & 1m54.0s & 100.00 \\
Poisson & Meta & BAL & 64 & 10 & 31.4s & 91.00 \\
Poisson & Levy-Baxter & BAL & 64 & 10 & 18m19.0s & 97.50 \\
\hline
\end{tabular}
}
\end{table}

\begin{table}[H] % Forces the table to appear in this exact location
\centering
\caption{Influence of numbers of neurons on commonly applied learning algorithms. The BP learning algorithm was applied.}
\label{table:merged_INFLUENCE_OF_NUMBERS_OF_NEURONS_BP}
\resizebox{0.9\linewidth}{!}{ % Adjusts the table width to fit the page
\begin{tabular}{|p{4cm}|p{4cm}|p{4cm}|p{4cm}|p{3cm}|p{3cm}|}
\hline
\textbf{Neuron model} & \textbf{Input data} & \textbf{Neurons number in each layer} & \textbf{Epochs number} & \textbf{Time} & \textbf{Accuracy [\%]} \\ [0.5ex] 
\hline\hline

LIF & Bernoulli & 16 & 10 & 14.2s & 97.00 \\
LIF & Bernoulli & 32 & 10 & 51.2s & 100.00 \\
LIF & Bernoulli & 64 & 10 & 3m19.0s & 99.00 \\
LIF & Bernoulli & 128 & 10 & 13m16.5s & 99.00 \\
LIF & Bernoulli & 512 & 10 & 223m53.5s & 99.50 \\
LIF & Bernoulli & 1024 & 10 & 729m36.0s & 100.00 \\
\hline
LIF & Markov & 16 & 10 & 28.7s & 95.92 \\
LIF & Markov & 32 & 10 & 32.4s & 97.42 \\
LIF & Markov & 64 & 10 & 49.9s & 88.22 \\
LIF & Markov & 128 & 10 & 1m51.0s & 97.64 \\
LIF & Markov & 512 & 10 & 26m9.0s & 96.08 \\
LIF & Markov & 1024 & 10 & 720m4.8s & 100.00 \\
\hline
LIF & Poisson & 16 & 10 & 2.9s & 69.50 \\
LIF & Poisson & 32 & 10 & 7.1s & 73.50 \\
LIF & Poisson & 64 & 10 & 22.3s & 47.00 \\
LIF & Poisson & 128 & 10 & 1m15.2s & 49.00 \\
LIF & Poisson & 512 & 10 & 24m48.0s & 49.00 \\
LIF & Poisson & 1024 & 10 & 84m10.5s & 49.00 \\
\hline\hline

Levy-Baxter & Bernoulli & 16 & 10 & 20.0s & 100.00 \\
Levy-Baxter & Bernoulli & 32 & 10 & 1m7.1s & 100.00 \\
Levy-Baxter & Bernoulli & 64 & 10 & 4m12.9s & 100.00 \\
Levy-Baxter & Bernoulli & 128 & 10 & 17m34.1s & 100.00 \\
Levy-Baxter & Bernoulli & 512 & 10 & 326m32.5s & 100.00 \\
Levy-Baxter & Bernoulli & 1024 & 10 & 79m48.4s & 49.00 \\
\hline
Levy-Baxter & Markov & 16 & 10 & 40.6s & 68.29 \\
Levy-Baxter & Markov & 32 & 10 & 1m2.3s & 98.05 \\
Levy-Baxter & Markov & 64 & 10 & 2m24.2s & 99.62 \\
Levy-Baxter & Markov & 128 & 10 & 7m51.3s & 99.88 \\
Levy-Baxter & Markov & 512 & 10 & 143m9.1s & 100.00 \\
Levy-Baxter & Markov & 1024 & 10 & 470m1.1s & 49.62 \\
\hline
Levy-Baxter & Poisson & 16 & 10 & 14.6s & 88.00 \\
Levy-Baxter & Poisson & 32 & 10 & 51.0s & 87.50 \\
Levy-Baxter & Poisson & 64 & 10 & 3m0.1s & 97.00 \\
Levy-Baxter & Poisson & 128 & 10 & 11m51.5s & 99.00 \\
Levy-Baxter & Poisson & 512 & 10 & 19m45.6s & 49.00 \\
Levy-Baxter & Poisson & 1024 & 10 & 210m11.0s & 49.00 \\
\hline
\end{tabular}
}
\end{table}

\par
Also taking into account the results obtained in the \cite{Paprocki2024} paper, namely that a large number of neurons in the network does not necessarily lead to significant improvements in transmission efficiency but can enhance the reliability of the system, we examined the influence of the number of neurons in individual layers on accuracy and computation time. In Tables \ref{table:merged_INFLUENCE_OF_NUMBERS_OF_NEURONS_BP}, \ref{table_combine_LiF_neuron}, \ref{combined_table_LB}, the influence of numbers of neurons on learning algorithms, taking account BP algorithms and bio-inspired learning algorithms were presented. The neural architecture made by LIF and LB neuron model, respectively, were widely investigated. Variants of neural networks that had 16, 32, 64, 128, 512 and 1024 nuerons in each layer were tested, respectively. In the case of BP learning algorithm (see, Table \ref{table:merged_INFLUENCE_OF_NUMBERS_OF_NEURONS_BP}) all computation was provided in the 10 epochs. Both networks composed of LIF and LB neurons achieved high accuracy in the case of data from Bernoulli and Markov processes, however, in the case of the Poisson process, the network model based on LIF neurons achieved a maximum accuracy of 73.50 percent with the number of neurons in the layer also 32. Then, as the number of neurons in the layers increased, the accuracy dropped below 50.00\%. The network based on the Levy-Baxter neuron model achieved an accuracy of more than 97. 00\% at 64, but the computation time was longer than when using the neural network based on the LIF model. In other cases, there is a visible trend towards an increase in precision as the number of neurons in the layers increases.

\begin{table}[H] % Forces the table to appear in this exact location
\centering
\caption{\scriptsize Influence of numbers of neurons on bio-inspired learning algorithms. The LIF neuron model was applied.} % Adjusted to a smaller size for consistency
\label{table_combine_LiF_neuron} % Modified label to avoid spaces
\resizebox{0.9\linewidth}{!}{ % Adjusting the width of the table to fit the page
\scriptsize % Reducing text size
\begin{tabular}{|p{2cm}|p{3cm}|p{5cm}|p{3cm}|p{3cm}|p{2cm}|}
\hline
\textbf{Input data} & \textbf{Learning algorithm} & \textbf{Neurons number in each layer} & \textbf{Epochs number} & \textbf{Time} & \textbf{Accuracy [\%]} \\ [0.5ex] 
\hline\hline

Bernoulli & Tempotron & 16 & 10 & 1.6s & 69.50 \\
Bernoulli & Tempotron & 32 & 10 & 2.5s & 66.50 \\
Bernoulli & Tempotron & 64 & 10 & 3.2s & 59.00 \\
Bernoulli & Tempotron & 128 & 10 & 7.02s & 96.00 \\
Bernoulli & Tempotron & 512 & 10 & 38.0s & 93.00 \\
Bernoulli & Tempotron & 1024 & 10 & 1m41.9s & 42.50 \\
\hline
Bernoulli & BAL & 16 & 10 & 2.7s & 55.50 \\
Bernoulli & BAL & 32 & 10 & 7.1s & 85.50 \\
Bernoulli & BAL & 64 & 10 & 22.2s & 86.00 \\
Bernoulli & BAL & 128 & 10 & 1m20.3s & 87.50 \\
Bernoulli & BAL & 512 & 10 & 21m21.6s & 95.00 \\
Bernoulli & BAL & 1024 & 10 & 82m27.9s & 97.00 \\
\hline
Bernoulli & STDP & 16 & 10 & 5.9s & 49.00 \\
Bernoulli & STDP & 32 & 10 & 16.9s & 75.00 \\
Bernoulli & STDP & 64 & 10 & 1m3.9s & 87.00 \\
Bernoulli & STDP & 128 & 10 & 4m5.3s & 91.50 \\
Bernoulli & STDP & 512 & 10 & 65m21.0s & 49.00 \\
Bernoulli & STDP & 1024 & 10 & 285m6.4s & 49.00 \\
\hline
Bernoulli & SDSP & 16 & 10 & 2.6s & 97.00 \\
Bernoulli & SDSP & 32 & 10 & 5.8s & 93.00 \\
Bernoulli & SDSP & 64 & 10 & 17.6s & 89.50 \\
Bernoulli & SDSP & 128 & 10 & 1m3.5s & 98.00 \\
Bernoulli & SDSP & 512 & 10 & 16m23.5s & 100.00 \\
Bernoulli & SDSP & 1024 & 10 & 58m36.4s & 100.00 \\
\hline\hline

Markov & Tempotron & 16 & 10 & 30.6s & 69.93 \\
Markov & Tempotron & 32 & 10 & 30.9s & 77.27 \\
Markov & Tempotron & 64 & 10 & 33.7s & 82.00 \\
Markov & Tempotron & 128 & 10 & 40.5s & 90.59 \\
Markov & Tempotron & 512 & 10 & 2m1.7s & 90.08 \\
Markov & Tempotron & 1024 & 10 & 4m33.1s & 90.08 \\
\hline
Markov & STDP & 16 & 10 & 49.0s & 49.00 \\
Markov & STDP & 32 & 10 & 14.9s & 75.00 \\
Markov & STDP & 64 & 10 & 57.5s & 87.00 \\
Markov & STDP & 128 & 10 & 3m45.1s & 91.50 \\
Markov & STDP & 512 & 10 & 59m33.4s & 53.00 \\
Markov & STDP & 1024 & 10 & 247m21.1s & 49.00 \\
\hline
Markov & BAL & 16 & 10 & 30.0s & 50.25 \\
Markov & BAL & 32 & 10 & 35.1s & 50.75 \\
Markov & BAL & 64 & 10 & 52.5s & 91.50 \\
Markov & BAL & 128 & 10 & 1m56.2s & 96.25 \\
Markov & BAL & 512 & 10 & 23m57.0s & 99.00 \\
Markov & BAL & 1024 & 10 & 93m53.3s & 99.75 \\
\hline
Markov & SDSP & 16 & 10 & 28.1s & 91.75 \\
Markov & SDSP & 32 & 10 & 31.4s & 97.00 \\
Markov & SDSP & 64 & 10 & 47.0s & 94.50 \\
Markov & SDSP & 128 & 10 & 1m46.4s & 92.25 \\
Markov & SDSP & 512 & 10 & 21m40.8s & 78.00 \\
Markov & SDSP & 1024 & 10 & 91m24.0s & 99.00 \\
\hline\hline

Poisson & Tempotron & 16 & 10 & 1.2s & 69.50 \\
Poisson & Tempotron & 32 & 10 & 1.9s & 68.00 \\
Poisson & Tempotron & 64 & 10 & 3.5s & 68.00 \\
Poisson & Tempotron & 128 & 10 & 7.3s & 89.50 \\
Poisson & Tempotron & 512 & 10 & 44.6s & 73.50 \\
Poisson & Tempotron & 1024 & 10 & 2m6.0s & 67.50 \\
\hline
Poisson & BAL & 16 & 10 & 3.5s & 79.00 \\
Poisson & BAL & 32 & 10 & 7.6s & 82.00 \\
Poisson & BAL & 64 & 10 & 25.0s & 48.50 \\
Poisson & BAL & 128 & 10 & 1m36.3s & 94.50 \\
Poisson & BAL & 512 & 10 & 24m48.1s & 49.00 \\
Poisson & BAL & 1024 & 10 & 109m38.3s & 48.50 \\
\hline
Poisson & STDP & 16 & 10 & 3.2s & 80.50 \\
Poisson & STDP & 32 & 10 & 7.3s & 83.50 \\
Poisson & STDP & 64 & 10 & 24.1s & 89.50 \\
Poisson & STDP & 128 & 10 & 1m27.5s & 49.00 \\
Poisson & STDP & 512 & 10 & 22m8.5s & 49.00 \\
Poisson & STDP & 1024 & 10 & 95m35.3s & 49.00 \\
\hline
Poisson & SDSP & 16 & 10 & 3.4s & 68.00 \\
Poisson & SDSP & 32 & 10 & 7.7s & 74.00 \\
Poisson & SDSP & 64 & 10 & 23.4s & 88.00 \\
Poisson & SDSP & 128 & 10 & 1m26.6s & 49.00 \\
Poisson & SDSP & 512 & 10 & 33m17.0s & 56.00 \\
Poisson & SDSP & 1024 & 10 & 101m5.0s & 64.50 \\
\hline
\end{tabular}
}
\end{table}

\begin{table}[H] 
\centering
\caption{\scriptsize Influence of numbers of neurons on bio-inspired learning algorithms. The LB neuron model was applied.}
\label{combined_table_LB}
\resizebox{0.9\linewidth}{!}{
\scriptsize
\begin{tabular}{|p{2cm}|p{3cm}|p{5cm}|p{3cm}|p{3cm}|p{2cm}|}
\hline
\textbf{Input data} & \textbf{Learning algorithm} & \textbf{Neurons number in each layer} & \textbf{Epochs number} & \textbf{Time} & \textbf{Accuracy [\%]} \\ [0.5ex] 
\hline\hline

Bernoulli & Tempotron & 16  & 10 & 2.2s      & 49.00 \\
Bernoulli & Tempotron & 32  & 10 & 4.5s      & 49.00 \\
Bernoulli & Tempotron & 64  & 10 & 12.2s     & 49.00 \\
Bernoulli & Tempotron & 128 & 10 & 49.0s     & 91.50 \\
Bernoulli & Tempotron & 512 & 10 & 2m38.4s   & 99.50 \\
Bernoulli & Tempotron & 1024& 10 & 23m45.1s  & 53.00 \\
\hline
Bernoulli & SDSP      & 16  & 10 & 8.7s      & 49.00 \\
Bernoulli & SDSP      & 32  & 10 & 27.5s     & 100.00 \\
Bernoulli & SDSP      & 64  & 10 & 1m50.5s   & 95.00 \\
Bernoulli & SDSP      & 128 & 10 & 6m48.0s   & 99.00 \\
Bernoulli & SDSP      & 512 & 10 & 74m51.5s  & 100.00 \\
Bernoulli & SDSP      & 1024& 10 & 585m52.5s & 74.00 \\
\hline
Bernoulli & STDP      & 16  & 10 & 4.6s      & 49.00 \\
Bernoulli & STDP      & 32  & 10 & 15.6s     & 49.00 \\
Bernoulli & STDP      & 64  & 10 & 56.1s     & 59.10 \\
Bernoulli & STDP      & 128 & 10 & 18m27.7s  & 99.50 \\
Bernoulli & STDP      & 512 & 10 & 59m20.9s  & 100.00 \\
Bernoulli & STDP      & 1024& 10 & 325m51.4s & 49.00 \\
\hline
Bernoulli & BAL       & 16  & 10 & 6.6s      & 97.00 \\
Bernoulli & BAL       & 32  & 10 & 22.2s     & 100.00 \\
Bernoulli & BAL       & 64  & 10 & 1m23.5s   & 98.00 \\
Bernoulli & BAL       & 128 & 10 & 5m22.5s   & 95.00 \\
Bernoulli & BAL       & 512 & 10 & 85m5.5s   & 100.00 \\
Bernoulli & BAL       & 1024& 10 & 311m47.0s & 52.00 \\
\hline\hline

Markov    & Tempotron & 16  & 10 & 29.7s     & 90.00 \\
Markov    & Tempotron & 32  & 10 & 32.2s     & 50.00 \\
Markov    & Tempotron & 64  & 10 & 36.0s     & 93.75 \\
Markov    & Tempotron & 128 & 10 & 1m7.0s    & 97.75 \\
Markov    & Tempotron & 512 & 10 & 4m7.8s    & 49.50 \\
Markov    & Tempotron & 1024& 10 & 10m53.1s  & 50.00 \\
\hline
Markov    & SDSP      & 16  & 10 & 31.6s     & 80.50 \\
Markov    & SDSP      & 32  & 10 & 36.4s     & 88.25 \\
Markov    & SDSP      & 64  & 10 & 57.65s    & 88.89 \\
Markov    & SDSP      & 128 & 10 & 1m59.65s  & 93.00 \\
Markov    & SDSP      & 512 & 10 & 25m9.65s  & 50.00 \\
Markov    & SDSP      & 1024& 10 & 91m24.0s  & 99.00 \\
\hline
Markov    & STDP      & 16  & 10 & 27.8s     & 47.25 \\
Markov    & STDP      & 32  & 10 & 32.7s     & 49.75 \\
Markov    & STDP      & 64  & 10 & 48.4s     & 53.25 \\
Markov    & STDP      & 128 & 10 & 1m47.9s   & 80.75 \\
Markov    & STDP      & 512 & 10 & 21m44.3s  & 100.00 \\
Markov    & STDP      & 1024& 10 & 89m45.9s  & 99.50 \\
\hline
Markov    & BAL       & 16  & 10 & 33.4s     & 87.50 \\
Markov    & BAL       & 32  & 10 & 38.5s     & 79.75 \\
Markov    & BAL       & 64  & 10 & 57.1s     & 50.00 \\
Markov    & BAL       & 128 & 10 & 2m5.2s    & 50.75 \\
Markov    & BAL       & 512 & 10 & 12m12.4s  & 96.25 \\
Markov    & BAL       & 1024& 10 & 106m54.7s & 61.75 \\
\hline\hline

Poisson   & Tempotron & 16  & 10 & 1.8s      & 49.00 \\
Poisson   & Tempotron & 32  & 10 & 3.4s      & 49.00 \\
Poisson   & Tempotron & 64  & 10 & 5.8s      & 91.00 \\
Poisson   & Tempotron & 128 & 10 & 21.0s     & 49.00 \\
Poisson   & Tempotron & 512 & 10 & 1m50.7s   & 49.00 \\
Poisson   & Tempotron & 1024& 10 & 5m39.6s   & 91.50 \\
\hline
Poisson   & SDSP      & 16  & 10 & 11.7s     & 49.00 \\
Poisson   & SDSP      & 32  & 10 & 41.0s     & 93.50 \\
Poisson   & SDSP      & 64  & 10 & 2m49.9s   & 95.50 \\
Poisson   & SDSP      & 128 & 10 & 11m9.2s   & 95.50 \\
Poisson   & SDSP      & 512 & 10 & 190m53.1s & 49.00 \\
Poisson   & SDSP      & 1024& 10 & 311m48.0s & 49.00 \\
\hline
Poisson   & STDP      & 16  & 10 & 4.9s      & 49.00 \\
Poisson   & STDP      & 32  & 10 & 10.1s     & 49.00 \\
Poisson   & STDP      & 64  & 10 & 31.1s     & 77.50 \\
Poisson   & STDP      & 128 & 10 & 2m3.2s    & 95.00 \\
Poisson   & STDP      & 512 & 10 & 7m31.0s   & 94.50 \\
Poisson   & STDP      & 1024& 10 & 95m50.7s  & 49.00 \\
\hline
Poisson   & BAL       & 16  & 10 & 33.4s     & 76.00 \\
Poisson   & BAL       & 32  & 10 & 2m6.3s    & 94.00 \\
Poisson   & BAL       & 64  & 10 & 8m6.7s    & 97.50 \\
Poisson   & BAL       & 128 & 10 & 32m29.2s  & 97.00 \\
Poisson   & BAL       & 512 & 10 & 546m25.4s & 49.00 \\
Poisson   & BAL       & 1024& 10 & 845m38.7s & 49.00 \\
\hline
\end{tabular}
}
\end{table}

In Table \ref{table_combine_LiF_neuron} we consider the influence of numbers of neurons in neural networks, which consists of LIF neurons on bio-inspired learning algorithms. Calculations were performed for 10 epochs. In the case of Bernoulli process and temporton learning algorithm increasing the number of neurons in the range 16-64 does not result in a significant increase in accuracy. For 128 and 512 neurons in each layer, we obtain an accuracy of over 93.00\%, while for 1024 we obtained only 42.50\%. When we classify a two-state Markov process, we get such accuracy over 90.00\% in the case of 128, 512 and 1024 neurons in layers. However, for the number of neurons in layers 64 it reaches 82.00\%. In turn, this learning algorithm does not work for the Poisson process. Only an accuracy greater than 80.00\% is achieved for 128 neurons in each layer. The SDSP algorithm allows to achieve high accuracy regardless of the number of neurons in the layers when classifying Bernoulli and Markov processes, while in the case of the Poisson process only when each layer has 64 neurons, i.e. 88.00\%. In the case of the BAL algorithm accuracy increases as the number of neurons in layers increases. Surprisingly, the BAL and SDSP learning algorithms can achieve accuracy above 80.00\% only for 64 neurons in layers when classifying Poisson processes. The STPD algorithm gives high accuracy for networks composed of a larger number of neurons, however not exceed 128 neurons in each layer. In turn, the computation times for the tempotron, BAL and SDSP learning algorithms are similar for all data. However, in the case of the STDP algorithm, with a large number of neurons in the layers, the computation time is more than twice as high.
\par
In turn, in Table \ref{combined_table_LB} the results obtained for neural network that consist of LB neuron was shown. Calculations were performed for 10 epochs. It turned out that the use of the tempotron learning algorithm gave high accuracy for all considered data, while the neural network architecture had 128, and 512 neurons in each layer. When the number of neurons in the layers was higher, the accuracy dropped below 50.00\%. A similar situation occurred when the number of neurons in the layers was smaller. For example, in the case of tempotron learning rule, the small layer sizes (16, 32 neurons) may lack sufficient representational capacity to capture the temporal and probabilistic dependencies in Bernoulli, Markov, and Poisson processes. This results in underfitting, where the model cannot adequately learn the patterns in the data. On the other hand, large layer sizes like 512, 1024 neurons may overfit the data, particularly for simpler processes. Overfitting can occur when the model memorizes specific patterns instead of generalizing, leading to poor performance on test data. A similar situation is with other learning algorithms, only BAL is exception.

\section{Discussion}

The results clearly show that classification performance in SNNs depends strongly on network scale and the statistical structure of input data. Hyperparameters such as learning rate and firing threshold must be tuned to network size; uniform settings often lead to underfitting in small networks or instability in large ones. In high-dimensional architectures, the increased number of synaptic weights expands the parameter space, making optimization more sensitive to initialization and regularization.
\par
The choice of neuron model shapes both computational capacity and task suitability. Simple models like perceptrons are computationally efficient for basic tasks but cannot capture temporal dynamics or nonlinear boundaries without deeper architectures \cite{Parlos1994}. Their performance is especially limited for stochastic inputs such as Poisson processes, where spike variability does not align well with their linear decision structure \cite{Dutta2017}.

More biologically plausible spiking models, e.g., LIF neurons capture temporal coding through spiking dynamics and refractory periods, improving performance on temporal classification tasks. However, they simplify real neuron behavior by omitting channel and synapse dynamics. The Levy–Baxter model combines deterministic and stochastic dynamics, enabling it to represent irregular neural firing more realistically \cite{LevyBaxter2002, Paprocki2020}, at the expense of computational complexity. Metaneurons \cite{Cheng2023} aggregate neuron populations into efficient computational units, improving scalability for complex tasks but reducing biological interpretability.

\par
Learning algorithm choice is equally critical: supervised approaches (backpropagation, tempotron) offer high accuracy and fast convergence for well-labeled datasets, while unsupervised rules (STDP, SDSP) are advantageous for limited labels or latent structure discovery. Optimal performance often arises from tailoring the learning rule to both the neuron model and the input statistics.

\par
The input process type Bernoulli, Poisson, or Markov has a measurable effect on model–algorithm combinations. Smaller networks struggle with the temporal complexity of Poisson and Markov sequences, whereas larger networks risk overfitting simpler Bernoulli data. LB neurons combined with tempotron learning perform well for two-state Markov data, effectively capturing transition probabilities ($p_{10}$, $p_{01}$).
\par
It is worth nothing that a key novel aspect of this study is the application of Lempel–Ziv Complexity as a classification mechanism for SNN outputs. By quantifying the structural complexity of spike trains, the LZC approach provides a lightweight and interpretable decision tool that avoids additional classifier layers, aligning well with low-power, real-time neuromorphic applications.

\section{Conclusions}
This study provides a systematic evaluation of how neuron model selection, network size, and learning rule jointly affect SNN classification efficiency. A key contribution is the introduction of a complexity-based classification method that uses Lempel-Ziv Complexity to analyze the structural regularity of spike trains. This method serves as a lightweight, interpretable decision mechanism, replacing conventional output layers and showing strong performance for noisy or temporally irregular signals. Our results show that the interaction between neuron model, network size, and learning algorithm is critical to classification accuracy. Moreover, data characteristics dictate the most effective model-algorithm combination, with LB neurons and tempotron learning particularly effective for temporally complex data. Also, the LZC-based decision approach complements biologically inspired models by enabling robust, low-power classification, opening paths for neuromorphic applications in biosignal processing. Future work will extend this framework to real biosignal datasets, explore adaptive LZC thresholds, and investigate hardware implementation for real-time SNN deployment.

\section*{Author contributions}
All authors contributed to the conception and design of the study. Material preparation, data collection, and analysis were performed by all authors. The first draft of the manuscript was written by all authors commented on previous versions of the manuscript. All authors read and approved the final manuscript.

\section*{Funding}
Not applicable.

\section*{Data Availability Statement}
Not applicable.

\end{document}